# Estimation de la résistance au cisaillement des sols granulaires hétérogènes à matrice

## Estimation of the shear strength of heterogeneous granular soils with matrix

N.M. Kouakou
*LEMTA (UMR 7563, Université de Lorraine – CNRS), Nancy, France*
*Bouygues Travaux Publics, Guyancourt, France*

O. Cuisinier, F. Masrouri
*LEMTA (UMR 7563, Université de Lorraine – CNRS), Nancy, France*

E. Lavallée, T. Le Borgne
*Bouygues Travaux Publics, Guyancourt, France*

**RÉSUMÉ :** La détermination des propriétés mécaniques des sols contenant des particules plus grandes que la taille admissible par les appareillages courants de laboratoire est problématique. Il est en effet nécessaire d'éliminer la fraction la plus grossière pour réaliser les essais, ce qui pose un problème de représentativité des résultats à l'échelle de l'ouvrage. Les méthodes de granulométrie parallèle et d'écrêtement se présentent comme un moyen d'estimer la résistance au cisaillement des granulaires hétérogènes à partir d'essais sur leur fraction fine. Cette étude se focalise sur l'effet encore méconnu du pourcentage de fines sur l'estimation par ces méthodes. Les résultats montrent que la méthode granulométrie parallèle permet de prédire l'angle de frottement du matériau initial malgré un pourcentage plus élevé de fines. En revanche, la cohésion du matériau initial est surestimée. Quant à la méthode d'écrêtement proposée par Fragaszy et al., le pourcentage de fines plus élevé dans le matériau modèle aboutit à une surestimation de la cohésion et une sous-estimation de l'angle de frottement du matériau initial.

**ABSTRACT:** The determination of the mechanical properties of soils containing particles larger than the allowable size by standard laboratory equipment is problematic. It is indeed necessary to eliminate the coarsest fraction to carry out the tests, which poses a problem of representativeness of the results at the scale of the structure. Parallel gradation and scalping methods are presented as a means of estimating the shear strength of heterogeneous granular soils from tests on their finer fraction. This study focuses on the effect of fines content on the estimation by these methods, which is not well known. The results showed that the parallel gradation method can predict the angle of friction of the initial material despite a high percentage of fines. However, the cohesion of the initial material is overestimated. As for the scalping method proposed by Fragaszy et al., the higher fines content in the modelled material resulted in an overestimation of the cohesion and an underestimation of the friction angle of the initial material.

**Mots-clés :** sol granulaire hétérogène ; résistance au cisaillement ; granulométrie parallèle ; écrêtement

# 1 INDRODUCTION

Les sols granulaires hétérogènes à matrice sont des matériaux souvent utilisés dans la construction des digues et des remblais. Ils sont constitués d'une partie fine (argileuse ou sableuse) et d'une partie grossière (graviers ou blocs rocheux). Dans cette étude, nous nous intéressons aux sols contenant des grains de diamètre supérieur à la taille admissible par les dispositifs courants de laboratoire ($d_{max} > 10$ mm). Leur utilisation dans des projets géotechniques requiert la détermination en laboratoire de leurs caractéristiques mécaniques, ce qui représente un enjeu particulier en raison de la présence des grains de grande taille. Une solution est alors d'employer des dispositifs de grandes dimensions. Par exemple, Estaire et Olalla (2005) utilisent une boîte de cisaillement direct de 1 m$^3$ de volume pour déterminer les paramètres de rupture des granulats de carrière et Ovalle et al. (2014) utilisent une cellule triaxiale de 1 m de diamètre et 1,5 m de hauteur. Cependant, ces appareillages, qui permettent de tester des sols de taille maximale de grain 160 mm, sont complexes à mettre en œuvre et inadaptés aux matériaux comportant des grains de taille plus grande.

Dès lors, certains auteurs ont cherché à développer une méthode pour estimer les paramètres de rupture de ces sols à partir d'essais sur leur fraction compatible avec les appareillages de cisaillement courants de laboratoire. Trois approches de reconstitution granulométriques ont été proposées dans la bibliographie : la substitution, l'écrêtement, et la granulométrie parallèle.

La substitution consiste à remplacer, à masse totale égale, les grains du sol initial dont la taille est supérieure à la taille admissible de l'appareillage par des grains de taille appropriée. Cette approche est peu utilisée parce qu'elle peut significativement modifier la courbe granulométrique du sol initial. Selon Vallé (2001), qui a réalisé son étude sur un sol naturel contenant des fines, la substitution conduit à une sous-estimation de la résistance mécanique du sol initial. Pedro (2004) conclut, à partir d'une étude sur un mélange sablo-graveleux, que cette approche donne une estimation proche de la résistance mécanique du sol initial. Ces conclusions contradictoires pourraient s'expliquer par la nature du matériau étudié et l'écart entre la courbe granulométrique du sol initial et celle du sol modèle.

L'écrêtement consiste à retirer du sol initial tous les grains de taille supérieure à la taille admissible par l'appareillage utilisé et à réaliser l'essai sur la fraction restante. Cette approche est peu utilisée également car elle sous-estime la résistance au cisaillement du sol initial selon Seif El Dine (2007) et conduit à une surestimation de la cohésion et une sous-estimation de l'angle de frottement du matériau initial fonction du pourcentage de grains écrêtés (Zhang et al., 2016). Quelques études ont montré l'importance de la densité sèche du sol écrêté dans l'utilisation de cette approche (Fragaszy et al., 1992; Pedro, 2004). Ainsi, lorsque le sol écrêté est testé à la même densité sèche que le sol initial, sa résistance au cisaillement est supérieure. En revanche, lorsqu'il est testé à sa densité sèche équivalente dans le sol initial, sa résistance au cisaillement est plus faible que celle du sol initial. Fragaszy et al. (1992) ont proposé une expression permettant de déterminer la densité sèche à laquelle tester le sol écrêté pour avoir une estimation satisfaisante de la résistance au cisaillement. Cette méthode, testée sur un sol alluvionnaire comportant 5% de fines a conduit à des résultats satisfaisants. Mais aucun auteur ne s'est intéressé à l'effet d'un pourcentage plus important de fines dans le matériau initial sur l'estimation par cette méthode.

L'approche la plus fréquemment utilisée est la granulométrie parallèle. Elle consiste à reconstituer, à partir des grains du sol de taille admissible par l'appareillage utilisé, un matériau de distribution granulométrique parallèle au sol initial. En utilisant cette méthode, Marachi et al. (1972) obtiennent un angle de frottement identique au matériau initial pour un même taux de rupture des grains. En revanche, Varadarajan et al. (2003) concluent à une sous-estimation de la résistance au cisaillement du sol initial lorsque les grains sont arrondis et une surestimation pour des grains anguleux. Ils expliquent cette différence de comportement par une rupture des grains moins importante pour le matériau ayant des grains arrondis. Par ailleurs, la plupart des études ont été réalisées sur des matériaux à faible pourcentage de fines de sorte que des interrogations demeurent sur l'effet d'un pourcentage important de fines dans le matériau modèle du fait de sa différence de nature par rapport au matériau initial. Ainsi, certains auteurs proposent d'attribuer au matériau modèle le même pourcentage de fines que le matériau initial mais les conclusions de Verdugo et De La Hoz (2006) montrent que cela sous-estime la résistance au cisaillement du sol initial. Ainsi, même si des résultats prometteurs ont été obtenus par la méthode de Fragaszy et la granulométrie parallèle sur les sols granulaires hétérogènes à matrice, l'effet d'un pourcentage important de fines dans le matériau initial reste méconnu.

Dans ce contexte, cette étude se focalise sur l'estimation de la résistance au cisaillement des sols granulaires hétérogènes ayant un pourcentage important de fines ( > 10%) par les méthodes d'écrêtement et de granulométrie parallèle. La méthode de substitution a été écartée car elle sous-estime la résistance au cisaillement de ces sols (Vallé, 2001). Ainsi, il s'agira de savoir si la méthode d'écrêtement proposée par Fragaszy et al. et la granulométrie parallèle permettent d'avoir une bonne estimation de la résistance au cisaillement malgré un pourcentage important de fines dans le matériau. Pour ce faire, les critères de rupture d'un matériau initial ayant un pourcentage élevé de fines et de deux matériaux modèles (obtenus par écrêtement et par granulométrie parallèle) seront déterminés puis comparés.

## 2 MATERIEL ET METHODES

### 2.1 Caractéristiques du matériau support de l'étude

Une grave naturelle villafranchienne de granulométrie 0-120 mm a été choisie comme matériau support de l'étude. Selon la classification GTR (2000), c'est un matériau de classe $C_1B_4$. Les dimensions des appareillages disponibles (voir section 2.2) imposant une taille maximale de grain de 30 mm, les grains de taille supérieure à 30 mm sont écartés. Le matériau 0-30 mm de granulométrie parallèle au matériau initial a été choisi comme matériau de base pour les essais de cisaillement. Son poids volumique sec maximal déterminé à partir d'un essai Proctor est de 21,1 kN/m$^3$ pour une teneur en eau correspondante de 8,3 %. À partir du matériau 0-30 mm, deux autres matériaux ont été reconstitués : un matériau 0-5 mm de granulométrie parallèle à celle de 0-30 mm et un matériau de granulométrie 0-15 mm obtenu par écrêtement du matériau 0-30 mm. Les courbes granulométriques des différents matériaux sont présentées sur la Courbes granulométriques du matériau initial et des matériaux modèlesFigure 1.

### 2.2 Appareillages de cisaillement

Trois boîtes de cisaillement direct à la boîte ont été utilisées : une petite boîte de dimensions 60 x 60 x 45 mm$^3$, une moyenne boîte de dimensions 150 x 150 x 180 mm$^3$ et une grande boîte de dimensions 300 x 300 x 180 mm$^3$.

La petite boîte est une boîte classique de cisaillement direct de mécanique des sols.

La grande boîte de cisaillement direct est constituée de deux demi-boîtes : la demi-boîte inférieure est fixe tandis que la demi-boîte supérieure est entraînée par un moteur à vitesse constante. La section de la grande boîte peut être réduite à 150 x 150 mm$^2$ grâce à des adaptateurs pour obtenir la moyenne boîte.

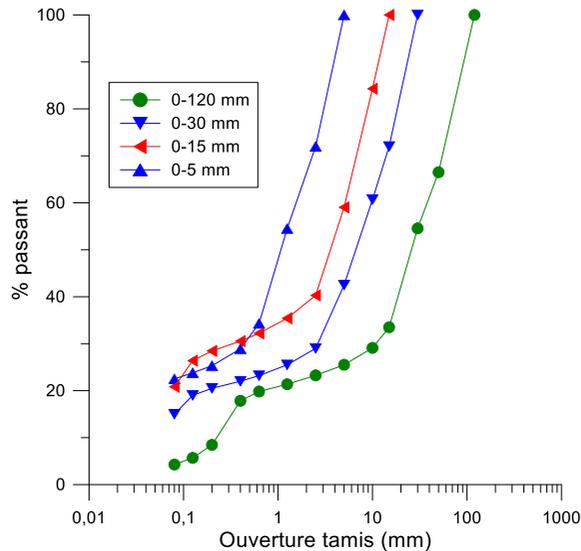

*Figure 1: Courbes granulométriques du matériau initial et des matériaux modèles*

Un système constitué d'un bâti et d'un piston hydraulique de capacité maximale 100 kN permet d'exercer un effort normal constant pendant l'essai. Des capteurs de déplacement de type LVDT sont installés afin de suivre les mouvements horizontaux et verticaux de l'éprouvette. Ces capteurs ont une course de 50 mm avec une précision de 0,2%. L'effort de cisaillement est mesuré par un capteur de force de capacité 100 kN. L'effort vertical est mesuré par un capteur de force placé entre le bâti et la plaque supérieure. Un capteur d'inclinaison installé sur la plaque supérieure permet de suivre les mouvements d'inclinaison de la plaque supérieure en cours d'essai. Les capteurs sont reliés à une centrale d'acquisition qui transmet les données à un logiciel pour enregistrement et visualisation.

## 2.3 Stratégie expérimentale

Le but des essais est d'étudier l'effet d'un pourcentage important de fines dans le matériau initial sur l'estimation par la méthode de granulométrie parallèle et la méthode d'écrêtement de Fragaszy et al. (1992).

Concernant la méthode de granulométrie parallèle, les matériaux modèles 0-30 (15% de fines) et 0-5 mm (22,5 % de fines) de distribution granulométrique parallèle au matériau initial ont été testés à la même masse volumique sèche de 2 Mg/m$^3$. Les critères de rupture de ces matériaux ont été déterminés à partir d'essais consolidés et drainés puis comparés.

Pour tester la méthode de Fragaszy et al., l'éprouvette de granulométrie 0-15 mm (21% de fines) obtenue par écrêtement de l'éprouvette 0-30 mm est reconstituée. Sa masse volumique sèche est calculée à partir de celle de l'éprouvette 0-30 mm et du pourcentage massique des grains écrêtés par l'équation 1 (Fragaszy et al., 1990) :

$$\rho_{0-15} = \frac{1-f}{\frac{1}{\rho_{0-30}} - \frac{f}{\alpha * \rho_s}} \quad (1)$$

$\rho_{0-15}$ (Mg/m$^3$): masse volumique sèche de l'éprouvette 0-15 mm ;
$\rho_{0-30}$ (Mg/m$^3$): masse volumique sèche de l'éprouvette 0-30 mm ;
$\rho_s$ (Mg/m$^3$) : masse volumique solide des grains écrêtés égale à 2,60 Mg/m$^3$ ;
$f$ (%) : pourcentage massique des grains écrêtés.
α : facteur tenant compte du supplément de vides dû à la présence des inclusions.

Les grains écrêtés représentant 28% du matériau 0-30 mm et le facteur α évalué à 0,903, la masse volumique sèche de l'éprouvette 0-15 mm est fixée à 1,89 Mg/m$^3$.

### 2.4 Préparation des éprouvettes

Afin de reconstituer la granulométrie voulue pour les éprouvettes, la fraction 0-30 mm de l'échantillon reçu a été séparée en différentes fractions granulométriques par tamisage :     < 0,08 ; 0,08-0,4 ; 0,4-0,63 ; 0,63-1,25 ; 1,25-2,5 ; 2,5-5 ; 5-10 ; 10-15 et 15-30 mm. Les différentes fractions sont mélangées selon les proportions données par la courbe granulométrique pour la confection des éprouvettes. Les matériaux sont humidifiés à la teneur en eau optimale (8,3%) et conservés dans des sacs hermétiques pendant au moins une nuit pour permettre une bonne homogénéité de l'eau dans l'échantillon. Les éprouvettes sont ensuite compactées statiquement en trois couches dans la moyenne et la grande boîte de cisaillement et en une couche dans la petite boîte jusqu'à atteindre la densité souhaitée. La saturation des éprouvettes se fait en remplissant d'eau déminéralisée le bac de l'appareillage de cisaillement et en laissant reposer pendant 3h minimum. Après la saturation, les éprouvettes sont consolidées sous l'effort normal voulu (correspondant aux contraintes normales de 50, 100 ou 200 kPa) puis cisaillées à une vitesse constante de 0,05 mm/min.

## 3   RÉSULTATS

### 3.1   Courbes de cisaillement des matériaux à granulométrie parallèle

Les résultats des essais de cisaillement des matériaux de granulométrie parallèle 0-5 et 0-30 mm sont présentés sur la Figure 2. Il apparait que les courbes des deux matériaux ne sont pas superposables. La courbe de cisaillement admet des valeurs au pic et résiduelle bien distinctes dans le cas du matériau 0-5 mm. Dans le cas du matériau 0-30 mm, il n'y a pas de pic apparent alors que les deux matériaux sont testés à la même densité. L'absence de pic pour le matériau 0-30 mm pourrait s'expliquer par les grandes dimensions de la boîte de cisaillement utilisée, ce qui rejoint les observations de Gotteland et al. (2000) et Bareither et al. (2008). En effet, ces auteurs ont constaté que lorsque les dimensions de la boîte de cisaillement augmentent, l'écart entre la résistance au pic et la résistance résiduelle se réduit jusqu'à disparaître complètement.

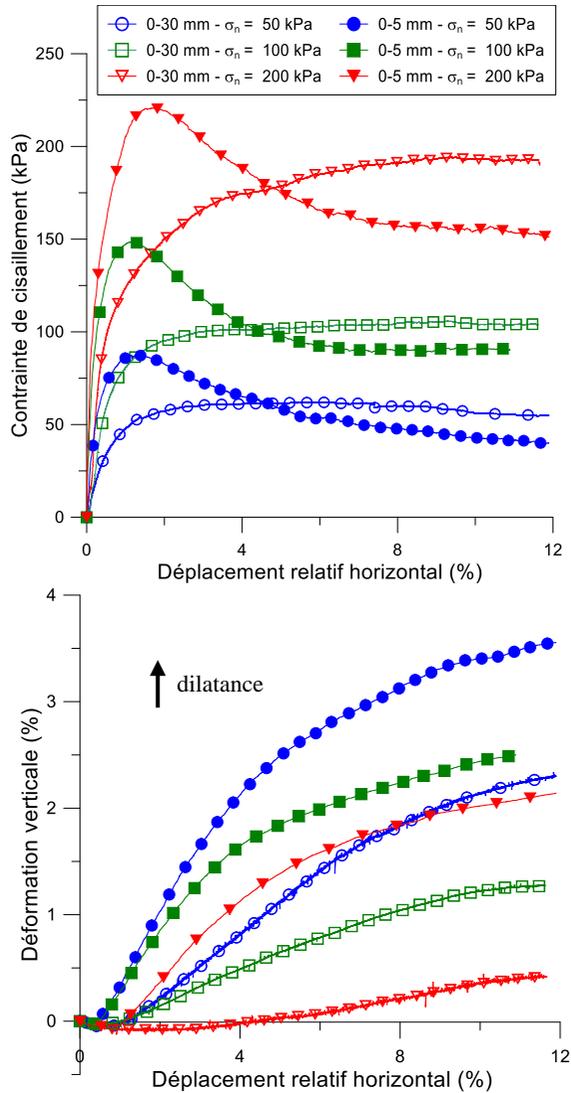

*Figure 2 : Courbes de cisaillement et de variation de volume des éprouvettes 0-5 et 0-30 mm*

Les courbes de variation de volume débutent par une phase de contractance suivie d'une phase de dilatance caractéristiques de l'état surconsolidé des éprouvettes dans les deux boîtes. La contractance du matériau 0-5 mm est très faible quelle que soit la contrainte normale avec une valeur maximale de 0,05% alors qu'elle intervient sur un intervalle de déformation plus long et avec une valeur plus élevée pour le matériau 0-30 mm pour les contraintes normales de 100 et 200 kPa. À l'opposé, la dilatance a une valeur plus élevée pour le matériau 0-5 mm comparativement au matériau 0-30 mm.
La méthode ne permet pas de reproduire les courbes de cisaillement et de variation de volume du matériau initial.

## *3.2* Courbes de cisaillement du matériau complet et du matériau écrêté

Les courbes de cisaillement du matériau complet 0-30 mm et du matériau écrêté 0-15 mm sous contraintes normales fixes de 50, 100 et 200 kPa sont présentées sur la Figure 3. L'écart entre les courbes des deux matériaux est de quelques pourcents sous les contraintes normales de 50 et 100 kPa. Les valeurs maximales de résistance au cisaillement sont quasi identiques. En revanche, le comportement des deux matériaux diffère sous une contrainte normale de 200 kPa. Le cisaillement est mobilisé plus progressivement pour le matériau écrêté et sa valeur maximale est inférieure à celle du matériau initial.

Les courbes de variation de volume montrent une contractance pour le matériau écrêté contrairement au matériau complet qui admet une phase de contractance suivie d'une phase de dilatance. La méthode ne permet pas de reproduire le comportement volumétrique du matériau initial.

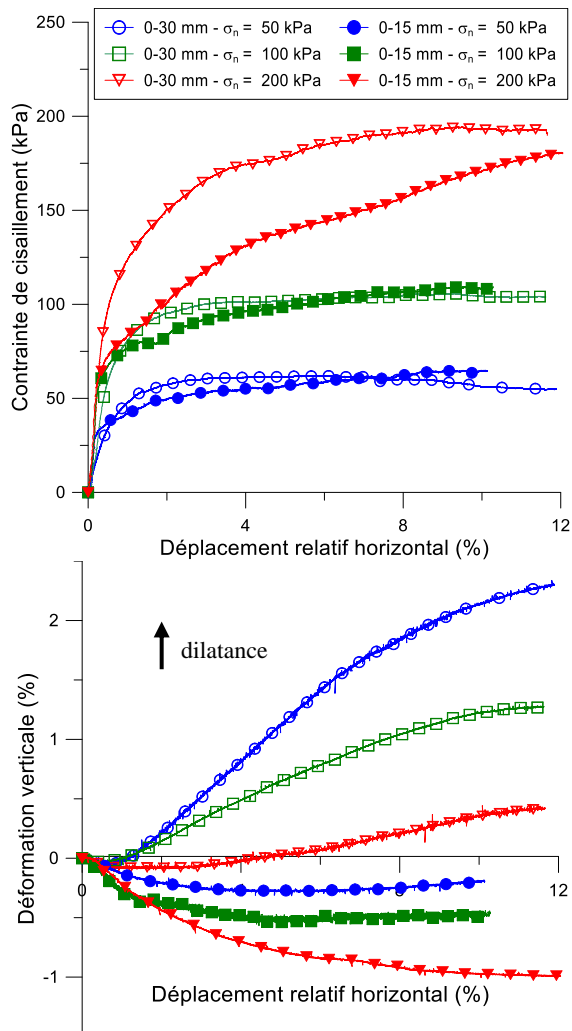

*Figure 3 : Courbes de cisaillement et de variation de volume des éprouvettes 0-15 et 0-30 mm*

## 3.3 Critères de rupture des matériaux de granulométrie parallèle

Les critères de rupture des matériaux à granulométrie parallèle selon le modèle de Mohr-Coulomb sont représentés sur la Figure 4. Il apparait que l'angle de frottement maximal des deux matériaux est identique. L'approche permet donc de prédire l'angle de frottement des matériaux grossiers dans la gamme des contraintes appliquées. L'important pourcentage de fines dans le matériau étudié n'a pas eu d'effet sur l'estimation de l'angle de frottement par la méthode.

En revanche, l'augmentation du pourcentage de fines dans le matériau 0-5 mm a abouti à une surestimation de la cohésion du sol initial. En effet, les fines dans le matériau servent de lien entre les grains plus gros ; ainsi l'augmentation de leur pourcentage contribue à améliorer la cohésion du sol.

Au vu des résultats obtenus, et en considérant le fait que les matériaux 0-5 mm et 0-30 mm sont de granulométrie parallèle au matériau 0-120 mm, une estimation des paramètres de rupture du matériau 0-120 mm peut être proposée. Ainsi, un angle de frottement de 41° et une cohésion nulle peuvent être attribués au matériau 0-120 mm car il contient un faible pourcentage de fines (4,3%).

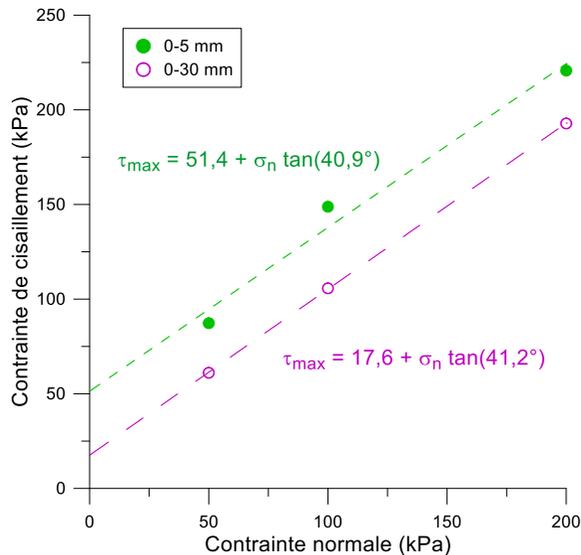

Figure 4 : Critères de rupture des matériaux de granulométrie parallèle 0-5 et 0-30 mm

## 3.4 Estimation par écrêtement suivant la méthode de Fragaszy et al. (1992)

La comparaison des critères de rupture du matériau complet 0-30 mm et de sa fraction écrêtée 0-15 mm (Figure 5) montrent une augmentation de la cohésion de 12 kPa et une diminution de l'angle de frottement de 4° avec l'écrêtement. L'utilisation de la densité proposée par Fragaszy et al. ne permet pas de prédire la cohésion et l'angle de frottement du matériau initial pour ce matériau ayant un important pourcentage de fines comparativement à celui étudié par Fragaszy et al. (15% contre 5%). Par ailleurs, les résistances au cisaillement des deux matériaux sont quasi-identiques sous les contraintes de 50 et 100 kPa. La méthode permettrait donc d'estimer correctement la résistance au cisaillement du matériau complet sous de faibles contraintes normales. En effet, les résultats présentés par Fragaszy et al. (1992) ont été obtenus sous des contraintes de confinement de 75 et 150 kPa. Ce constat pourrait s'expliquer par le fait que sous de faibles contraintes normales, la cohésion est un terme prépondérant de la résistance au cisaillement du sol. En effet, la cohésion plus élevée du sol écrêté et son angle de frottement plus faible font que pour une gamme de contrainte normale (entre 50 et 150 kPa dans cette étude), les résistances au cisaillement des deux matériaux sont quasi identiques.

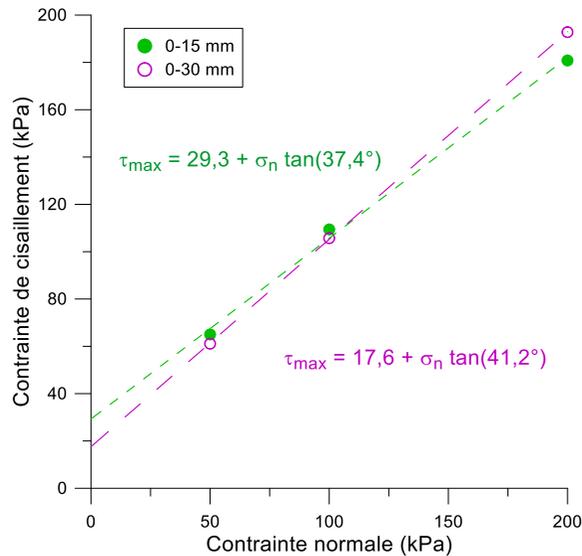

*Figure 5 : Critères de rupture des matériaux 0-15 et 0-30 mm*

## 4 CONCLUSION

L'objectif de cette étude était de déterminer l'effet d'un pourcentage important de fines sur l'estimation de la résistance au cisaillement d'un sol granulaire hétérogène par les méthodes de granulométrie parallèle et d'écrêtement. Les conclusions de cette étude sont les suivantes :

La méthode de granulométrie parallèle permet d'estimer correctement l'angle de frottement des matériaux granulaires hétérogènes à matrice pour un important pourcentage de fines dans le matériau initial. Par contre, l'augmentation importante du pourcentage de fines conduit à une surestimation de la cohésion.

La méthode d'estimation par écrêtement proposée par Fragaszy et al. (1992) aboutit à une surestimation de la cohésion et une sous-estimation de l'angle de frottement pour les matériaux contenant un important pourcentage de fines. Néanmoins, elle permet d'estimer leur résistance au cisaillement sous de faibles contraintes normales.

La détermination du pourcentage de fines maximal dans le matériau modèle assurant une bonne estimation de l'angle de frottement par la méthode de granulométrie parallèle pourrait constituer une suite à ce travail.

## REMERCIEMENTS

Modelling and experimental assessment of geomaterials